\documentclass[12pt,a4paper]{article}

\usepackage{graphicx}
\usepackage{epsfig}
\usepackage{amssymb}
\usepackage{amsmath}
\usepackage{latexsym}

\newcommand{\capdef}{}
\newcommand{\mycaption}[2][\capdef]{\renewcommand{\capdef}{#2}%
       \caption[#1]{{\footnotesize #2}}}
\makeatletter
\renewcommand{\fnum@table}{\textbf{\tablename~\thetable}}
\renewcommand{\fnum@figure}{\textbf{\figurename~\thefigure}}
\makeatother

\makeatletter
\renewcommand{\section}{\@startsection{section}{1}{0em}{-\baselineskip}%
{\baselineskip}{\normalfont\large\bfseries}}
\renewcommand{\subsection}%
{\@startsection{subsection}{2}{0em}{-0.7\baselineskip}%
{0.7\baselineskip}{\normalfont\bfseries}}
\makeatother

\renewcommand{\baselinestretch}{1.12}

\newcommand{\MH}{M_{H^{++}}}

\newcommand{\beq}{\begin{equation}}
\newcommand{\eeq}{\end{equation}}

\newcommand{\bea}{\begin{eqnarray}}
\newcommand{\eea}{\end{eqnarray}}

\newcommand{\BRee}{\ensuremath{\text{BR}_{ee}}}
\newcommand{\BRem}{\ensuremath{\text{BR}_{e\mu}}}
\newcommand{\BRmm}{\ensuremath{\text{BR}_{\mu\mu}}}
\newcommand{\BRet}{\ensuremath{\text{BR}_{e\tau}}}
\newcommand{\BRmt}{\ensuremath{\text{BR}_{\mu\tau}}}

\begin{document}


\renewcommand{\thefootnote}{\alph{footnote}}

\begin{flushright}
CERN-PH-TH/2007-255\\
IFIC/07-75, FTUV-07-1210\\
\end{flushright}

\vspace*{0.5cm}

\renewcommand{\thefootnote}{\fnsymbol{footnote}}
\setcounter{footnote}{0}

\begin{center} 
\Large\textbf{
Neutrino mass hierarchy and Majorana CP phases within 
the Higgs triplet model at the LHC
}
\end{center}

\vspace*{0.5cm}

\begin{center} 
{\bf Julia Garayoa\footnote{email: \tt garayoa AT ific.uv.es} 
and Thomas Schwetz\footnote{email: \tt schwetz AT cern.ch}}

\vspace*{2mm}\it 
$^*$Departament de F\'{i}sica Te\`orica and IFIC, Universitat de Val\`encia-CSIC,
Edifici Instituts d'Investigaci\'o, Apt 22085, E-46071 Val\`encia, Spain\\
$^\dagger$CERN, Physics Department, Theory Division, CH-1211 Geneva 23, Switzerland
\end{center}

\vspace*{0.3cm}

\begin{abstract}
  Neutrino masses may be generated by the VEV of an $SU(2)_L$ Higgs
  triplet. We assume that the doubly charged component of such a
  triplet has a mass in the range of several 100~GeV, such that it is
  accessible at LHC. Its decay into like-sign leptons provides a clean
  experimental signature, which allows for a direct test of the
  neutrino mass matrix. By exploring the branching ratios of this decay
  into leptons of various flavours, we show that within this model the
  type of the neutrino mass spectrum (normal, inverted or
  quasi-degenerate) might actually be resolved at the
  LHC. Furthermore, we show that within the Higgs triplet model for
  neutrino mass the decays of the doubly charged scalar into like-sign
  lepton pairs at the LHC provide a possibility to determine the
  Majorana CP phases of the lepton mixing matrix.
\end{abstract}

\renewcommand{\thefootnote}{\arabic{footnote}}
\setcounter{footnote}{0}

\newpage

\vspace*{0.5cm}

\section{Introduction}

Recent developments in neutrino physics demand for an extension of the
Standard Model in order to give mass to the neutrinos. A popular way
to achieve this goal is to introduce right-handed singlet neutrinos.
An alternative, equally valid and rather economical possibility is to
extend the scalar sector of the Standard Model. In addition to the
Higgs doublet, scalar representations consistent with the
$SU(2)_L\times U(1)_Y$ gauge group and the possible fermionic
bilinears are~\cite{Konetschny:1977bn} a triplet, a singlet with
charge +1, or a singlet with charge +2, see
Refs.~\cite{Gelmini:1980re, Zee:1980ai, Zee:1985id} for corresponding
models. In this work we focus on the first mentioned possibility,
namely a scalar $SU(2)_L$ triplet. Such a triplet arises naturally in
many extensions of the Standard Model, for example in left-right
symmetric models~\cite{left-right}, or in Little Higgs
theories~\cite{littleH, Han:2005nk}. 
When the neutral component of the triplet acquires a vacuum expectation
value (VEV), $v_T$, a Majorana mass term for neutrinos is generated at
tree level, proportional to $v_T$. In order to obtain small neutrino
masses this VEV and/or the corresponding Yukawa couplings have to be
very small. If a very high energy scale $M \gg v = 246$~GeV is
associated to the triplet, one obtains the well-known seesaw
(type-II) relation $v_T \sim v^2/M$ as explanation for the smallness
of neutrino masses~\cite{type-II, Schechter:1980gr, Ma:1998dx}.

Here we consider a different scenario, assuming that the triplet
states have masses not too far from the electroweak scale. The Higgs
potential of the Standard Model Higgs doublet $\phi$ and the triplet
$\Delta$ contains a term $\mu \phi\Delta\phi$, which breaks lepton
number explicitly. Assuming that all other mass parameters in the
potential are of the electroweak scale $v$, the minimisation of the
potential leads to the relation for the triplet VEV $v_T \sim \mu$,
see e.g.~\cite{Grimus:1999fz}. The hierarchy $\mu \ll v$ may find an
explanation for example through extra dimensions~\cite{Ma:2000wpa}.
A Higgs triplet slightly below the TeV scale is the generic situation
in Little Higgs theories~\cite{littleH}, see Ref.~\cite{Han:2005nk}
for a discussion of neutrino masses in this framework.  Other examples
for models with TeV scale triplets responsible for neutrino masses can
be found, e.g., in Refs.~\cite{331, sahu, Dorsner:2007fy}. Our
phenomenological analysis does not rely on a specific model
realisation, apart from the assumption that neutrino masses arise from
a triplet with masses in the TeV range.

The hypothesis of such a Higgs triplet can be tested at collider
experiments. In particular, if kinematically accessible, the doubly
charged component of the triplet $H^{++}$ will be produced in high
energy collisions, and its decay into two equally charged leptons
provides a rather spectacular signature, basically free of any
Standard Model background. This process has been studied extensively
in the literature (see Refs.~\cite{Gunion:1989in, Huitu:1996su,
Gunion:1996pq, Dion:1998pw, Muhlleitner:2003me, Azuelos:2005uc,
Akeroyd:2005gt, Hektor:2007uu, Han:2007bk} for an incomplete list),
and has been used to look for doubly charged scalars at LEP~\cite{LEP}
and Tevatron~\cite{tevatron}. These searches resulted in lower bounds
for the mass of the order $\MH \gtrsim 130$~GeV. Therefore, we will
consider in the following masses in the range $130\,\text{GeV}
\lesssim \MH \lesssim 1\,\text{TeV}$, above the present bound but
still in reach for LHC.

If the Higgs triplet is responsible for the neutrino mass the decay
rate for $H^{++} \to \ell^+_a \ell^+_b$ is proportional to the modulus
of the corresponding element of the neutrino mass matrix
$|M_{ab}|^2$. This opens a phenomenologically very interesting link
between neutrino and collider physics\footnote{Such a link exists also
in other classes of models, see for example~\cite{Babu:2002uu,
AristizabalSierra:2003ix, Chen:2007dc, Porod:2000hv,
AristizabalSierra:2006ri, Bajc:2007zf}. However, in most cases the
connection between collider signals and the neutrino mass matrix is
much less direct as in the Higgs triplet model.}, and by the
observation of like-sign lepton events at LHC a direct test of the
neutrino mass matrix becomes possible.  
In this work we assume that a doubly charged Higgs is indeed
discovered at LHC, and we use the information from the decays
$H^{++}\to \ell^+_a \ell^+_b$ to learn something about neutrinos,
under the hypothesis that the neutrino mass matrix is dominantly
generated by the triplet VEV.  

Current neutrino data leave some ambiguities for the neutrino mass
spectrum. The neutrino mass states can be ordered normally or inverted,
and the masses can be hierarchical or quasi-degenerate. We will show
that under the above assumptions actually LHC might play a decisive
role in distinguishing these possibilities. Furthermore, we show that
it might be possible to determine the Majorana
phases~\cite{Schechter:1980gr, Bilenky:1980cx} in the lepton mixing
matrix, which in general is a very difficult task.
Implications of the different possibilities of the neutrino mass
spectrum for the decay of a doubly charged scalar in the Higgs triplet
model have been considered previously in Ref.~\cite{Chun:2003ej}, see
also \cite{Akeroyd:2005gt}. Building upon the results obtained there,
we perform a full parameter scan including all complex phases,
which---as we will see---play a crucial role for the relevant
observables.

The outline of the paper is as follows.  In Sec.~\ref{sec:framework}
we present the general framework, where in Sec.~\ref{sec:nu-masses} we
review how the neutrino mass matrix arises in the Higgs triplet model,
and in Sec.~\ref{sec:LHC} we discuss the signature of the model at
LHC. Sec.~\ref{sec:results} contains the main results of our work.
After describing our analysis in Sec.~\ref{sec:analysis}, we discuss
in Sec.~\ref{sec:branching} how the branching ratios of the doubly
charged scalar depend on the parameters of the neutrino mass
matrix. In Sec.~\ref{sec:spectrum} we investigate the possibility to
determine the type of the neutrino mass spectrum from like-sign lepton
events at LHC, whereas in Sec.~\ref{sec:phases} we show that within
this framework indeed Majorana phases can be determined. Concluding
remarks follow in Sec.~\ref{sec:conclusions}.

\section{Framework}
\label{sec:framework}

\subsection{Neutrino masses from a Higgs triplet}
\label{sec:nu-masses}

If an $SU(2)_L$ Higgs triplet with hypercharge $Y=2$ is present in the
theory the following renormalisable term appears in the Yukawa sector
of the Lagrangian:
\begin{equation}\label{eq:yukawa}
\mathcal{L}_\Delta = f_{ab}\, L^T_a C^{-1} \,i\tau_2 \Delta \, L_b  
+ \text{h.c.}\,,  
\end{equation}
where the indices $a,b = e,\mu,\tau$ label flavours, $L_a$ are the
lepton doublets, $C$ is the charge conjugation matrix, $\tau_2$ is the
Pauli matrix, $\Delta$ denotes the scalar triplet, and $f_{ab}$ is a
symmetric complex Yukawa matrix. Without loss of generality we work in
the mass basis of the charged leptons. The components of the triplet
are given by:
\begin{equation}
\Delta = \left(\begin{array}{cc}
H^+ / \sqrt{2} & H^{++} \\
H^0 & -H^+ / \sqrt{2} 
\end{array}\right) \,.
\end{equation}
The VEV of the neutral component $\langle H^0 \rangle \equiv v_T/\sqrt{2}$ 
induces a Majorana mass term for the neutrinos:
\begin{equation}\label{eq:Mnu}
\frac{1}{2} \nu_{La}^T C^{-1} M_{ab} \, \nu_{Lb} + \text{h.c.}
\qquad\text{with}\qquad
M_{ab} = \sqrt{2} \, v_T \, f_{ab} \,.
\end{equation}
We assume in the following that this is the sole source for neutrino
masses (or at least the dominant contribution). As usual the neutrino
mass matrix $M_{ab}$ is diagonalised by:
\begin{equation}\label{eq:diag}
M = U \text{diag}(m_1, m_2, m_3) U^T \,.
\end{equation}
For the PMNS matrix $U$ we adopt the parametrisation
\begin{equation}\label{eq:U}
\begin{aligned}
 U &= V \, \mathrm{diag}
 (e^{i\frac{\alpha_1}{2}},
  e^{i\frac{\alpha_2}{2}},
  e^{i\frac{\alpha_3}{2}}) \qquad \text{with} \\
 V &= \left(
  \begin{array}{ccc}
  c_{12}c_{13} & 
  s_{12}c_{13} & s_{13}e^{-i\delta}\\
  -c_{23}s_{12}-s_{13}s_{23}c_{12}e^{i\delta} &
  c_{23}c_{12}-s_{13}s_{23}s_{12}e^{i\delta}  &
  s_{23}c_{13}\\
  s_{23}s_{12}-s_{13}c_{23}c_{12}e^{i\delta} &
  -s_{23}c_{12}-s_{13}c_{23}s_{12}e^{i\delta} &
  c_{23}c_{13}
  \end{array}
  \right) \,
\end{aligned}
\end{equation}
where $\delta$ is the so-called Dirac CP violating phase which is in
principle measurable in neutrino oscillation experiments, and
$\alpha_i$ are the Majorana phases~\cite{Schechter:1980gr,
Bilenky:1980cx}. Note that only relative phases $\alpha_{ij}\equiv
\alpha_i - \alpha_j$ are physical, and therefore there are only two
independent Majorana phases. Neutrino oscillation data determine the
so-called solar and atmospheric oscillation
parameters~\cite{Maltoni:2004ei}:
\beq\label{eq:osc-params}
\begin{aligned}
  \sin^2\theta_{12} = 0.32\pm 0.023 &\,,\quad
  \Delta m^2_{21} = (7.6\pm 0.20)\times 10^{-5}\, \text{eV}^2 \,,\\
  \sin^2\theta_{23} = 0.50\pm 0.063 &\,,\quad
  |\Delta m^2_{31}| = (2.4\pm 0.15)\times 10^{-3}\, \text{eV}^2 \,,
\end{aligned}\eeq
where we give $1\sigma$ errors and $\Delta m^2_{ij} \equiv m^2_i -
m^2_j$. For the mixing angle $\theta_{13}$ there is only an upper
bound,
\begin{equation}\label{eq:th13}
\sin^2\theta_{13} < 0.05 \quad\text{at}~3\sigma\,,
\end{equation}
whereas nothing is known about the phases $\delta, \alpha_{ij}$. The
ordering of the mass states is determined by the sign of $\Delta
m^2_{31}$: for normal hierarchy (NH) $\Delta m^2_{31} > 0$, whereas
for inverted hierarchy (IH) we have $\Delta m^2_{31} < 0$. We denote
the lightest neutrino mass by $m_0$, hence,
\begin{equation}
m_0 = \left\{\begin{array}{l@{\qquad}l}
      m_1 & \text{(NH)} \\
      m_3 & \text{(IH)} \end{array}\right. \,.
\end{equation}
If $m_0 \gtrsim \sqrt{|\Delta m^2_{31}|} \simeq 0.05$~eV the neutrino
mass spectrum is quasi-degenerate (QD). The most stringent bound on
the absolute scale of the neutrino mass comes from cosmology, which is
sensitive to the sum of the three masses. In a recent analysis
\cite{Hannestad:2007tu} the upper bound $\sum_i m_i < 0.5$~eV at
95\%~CL has been obtained, which translates into $m_0 < 0.16
\,\text{eV}$.  Since this corresponds to the QD regime the bound is
the same for NH and IH. Taking into account Eq.~(\ref{eq:Mnu}), the
constraint from cosmology applies directly to the product of triplet
VEV and Yukawas:
\begin{equation}\label{eq:cosmo-bound}
v_T f_{ab} \lesssim 10^{-10} \,\text{GeV}\,.
\end{equation}

\subsection{Doubly charged scalars at the LHC}
\label{sec:LHC}

At the LHC the process 
\begin{equation}\label{eq:prod}
pp\to H^{++}H^{--} \to \ell^+\ell^+ \,\ell^-\ell^- 
\end{equation}
provides a very spectacular signature, namely two like-sign lepton
pairs with the same invariant mass and no missing transverse momentum,
which has essentially no Standard Model background. The pair
production of the doubly charged scalar occurs by the Drell-Yan
process $q \overline{q} \to \gamma^*, Z^* \to H^{--} H^{++}$, with a
sub-dominant contribution also from two-photon fusion $\gamma \gamma
\to H^{--} H^{++}$. The cross section is not suppressed by any small
quantity (such as the Yukawas or the triplet VEV) and depends only on
the mass $\MH$, see e.g.~\cite{Gunion:1996pq, Han:2007bk}. QCD
corrections at next-to-leading order have been
calculated~\cite{Muhlleitner:2003me}.\footnote{Let us note
that---depending on the mass splitting between the double and single
charged components of the triplet---also the channel
$q'\overline{q}\to H^{\pm\pm}H^\mp$ may significantly contribute to
the production of doubly charged scalars, see e.g.~\cite{Dion:1998pw,
Akeroyd:2005gt}.}
The cross section for $H^{--} H^{++}$ pair production at the LHC
ranges from 100~fb for a Higgs mass $\MH = 200$~GeV to 0.1~fb for $\MH
= 900$~GeV~\cite{Han:2007bk}. Hence, if the doubly charged scalar is
not too heavy a considerable number of them will be produced at LHC
assuming an integrated luminosity of order 100~fb$^{-1}$.

The rate for the decay $H^{++} \to \ell^+_a \ell^+_b$ is given by
\begin{equation}\label{eq:decay-rate}
\Gamma(H^{++} \to \ell^+_a \ell^+_b) = \frac{1}{4\pi(1+\delta_{ab})}\,
  |f_{ab}|^2 \MH \,,
\end{equation}
with $\delta_{ab} = 1 \,(0)$ for $a=b \,(a \neq b)$. Hence, the rate
is proportional to the corresponding element of the neutrino mass
matrix $|M_{ab}|^2$. This observation is the basis of our
analysis. Using Eqs.~(\ref{eq:Mnu}) and (\ref{eq:decay-rate}) the
branching ratio can be expressed as
\beq \label{BR}
  \text{BR}_{ab} \equiv
  \text{BR}(H^{++} \to \ell_a^+ \ell_b^+) \equiv
  \frac{\Gamma(H^{++} \to \ell^+_a \ell^+_b)}
  {\sum_{cd}\Gamma(H^{++} \to \ell^+_c \ell^+_d)} = 
  \frac{2}{(1+\delta_{ab})} \,
  \frac{|M_{ab}|^2}
  {\sum_{cd}|M_{cd}|^2} \,,
\eeq
and from Eq.~(\ref{eq:diag}) and the unitarity of $U$ follows
\beq
  \sum_{cd}|M_{cd}|^2 = \sum_{i=1}^3 m_i^2 = \left\{
  \begin{array}{l@{\qquad}l}
    3m_0^2 + \Delta m^2_{21} + \Delta m^2_{31} & \text{(NH)} \\
    3m_0^2 + \Delta m^2_{21} + 2|\Delta m^2_{31}| & \text{(IH)} 
  \end{array}\right. \,.
\eeq

In addition to the lepton
channel the doubly charged Higgs can in principle decay also into the
following two-body final states including singly charged Higges and/or
the $W$:
\begin{equation}\label{eq:decays}
H^{++} \to H^+ H^+ \,,\quad
H^{++} \to H^+ W^+ \,,\quad
H^{++} \to W^+ W^+ \,.
\end{equation}
The first two decay modes depend on the mass splitting within the
triplet. We assume in the following that they are kinematically
suppressed. The rate for the $WW$ mode is given by
\begin{equation}
\Gamma(H^{++} \to W^+ W^+) \approx \frac{v_T^2 \MH^3}{2\pi v^4} \,,
\end{equation}
where $v = 246$~GeV is the VEV of the Standard Model Higgs doublet,
and we have used $\MH \gg M_W$, see e.g., Ref.~\cite{Han:2007bk} for
full expressions and a discussion of possibilities to observe this
process at LHC. Hence, the branching ratio between $\ell^+\ell^+$ and
$W^+W^+$ decays is controlled by the relative magnitude of the triplet
Yukawas $f_{ab}$ and the VEV $v_T$. The requirement $ \Gamma(H^{++}
\to W^+ W^+) \lesssim \Gamma(H^{++} \to \ell^+_a \ell^+_b)$, together
with the constraint from Eq.~(\ref{eq:cosmo-bound}) implies:
\begin{equation}\label{eq:vev-bound}
\frac{v_T}{v} \lesssim 10^{-6} 
\left(\frac{100\,\text{GeV}}{\MH}\right)^{1/2} \,.
\end{equation}
The triplet VEV contributes to the $\rho$ parameter at tree level
as~\cite{Gelmini:1980re} $\rho \approx 1 - 2(v_T/v)^2$. The constraint
from electroweak precision data $\rho = 1.0002^{+0.0024}_{-0.0009}$ at
2$\sigma$~\cite{Yao:2006px} translates into $v_T/v < 0.02$, which is
savely satisfied by requiring Eq.~(\ref{eq:vev-bound}).

In this model contributions to lepton flavour violating processes,
$g_\mu - 2$, and in principle also to the electron electric dipole
moment are expected, see e.g.~\cite{Chun:2003ej, Cuypers:1996ia,
Kakizaki:2003jk} and references therein. Following
Refs.~\cite{Chun:2003ej, Kakizaki:2003jk}, the most stringent
constraint on the Yukawa couplings $f_{ab}$ comes from $\mu\to e e e$,
a process which occurs at tree level via Eq.~(\ref{eq:yukawa}). The
branching ratio for this decay is given by~\cite{Kakizaki:2003jk}:
\begin{equation}
  \text{BR}(\mu\to eee) = \frac{1}{4G_F^2}
    \frac{|f_{ee}^* f_{e\mu}|^2}{\MH^4} \approx
    20 \left( \frac{\MH}{100\,\text{GeV}} \right)^{-4} 
    |f_{ee}^* f_{e\mu}|^2\,.
\end{equation}    
Hence, the experimental bound BR$(\mu\to eee) <
10^{-12}$~\cite{Yao:2006px} constrains the combination $|f_{ee}^*
f_{e\mu}| \lesssim 2\times 10^{-7}(\MH/100\,\text{GeV})^2$.  
Assuming that all $f_{ab}$ have roughly the same order of magnitude we
obtain an estimate for the interesting range of the Yukawa couplings:
\begin{equation}\label{eq:yuk-range}
  4\times 10^{-7} \left(\frac{\MH}{100\,\text{GeV}}\right)^{1/2}
  \lesssim f_{ab} \lesssim
  5\times 10^{-4} \left(\frac{\MH}{100\,\text{GeV}}\right) \,,
\end{equation}
where the lower bound emerges from Eq.~(\ref{eq:vev-bound}) assuming
that the bound (\ref{eq:cosmo-bound}) is saturated. We see that
several orders of magnitude are available for the Yukawa
couplings. For $f_{ab}$ close to the lower bound of
Eq.~(\ref{eq:yuk-range}) the decay $H^{++} \to W^+W^+$ will become
observable at LHC, whereas close to the upper bound a signal in future
searches for lepton flavour violation is expected, where the details
depend on the structure of the neutrino mass matrix~\cite{Chun:2003ej,
Kakizaki:2003jk}. The interval for the Yukawas from
Eq.~(\ref{eq:yuk-range}) implies a triplet VEV roughly in the keV to
MeV range.

The basic assumption in our analysis is that a sufficient number of
like-sign leptons is observed. If some of the decay modes of
Eq.~(\ref{eq:decays}) are present the number of dilepton events will
be reduced according to the branching. If enough events from both types
of decay (leptonic and non-leptonic) were observed in principle an
order of magnitude estimate for the Yukawa couplings $f_{ab}$ and the
triplet VEV $v_T$ might be possible~\cite{Gunion:1996pq,
Akeroyd:2005gt}. Here we do not consider this case and use only
dilepton events, and therefore, we do not obtain any information on
the overall scale of the $f_{ab}$ in addition to
Eq.~(\ref{eq:yuk-range}). 

\section{Numerical analysis and results}
\label{sec:results}

\subsection{Description of the analysis}
\label{sec:analysis}

As mentioned above, we focus in our analysis on the process
(\ref{eq:prod}), which provides the clean signal of four leptons,
where the like-sign lepton pairs have the same invariant mass, namely
the mass of the doubly charged Higgs. Given the fact that the
branching $H^{++} \to \ell_a^+ \ell_b^+$ is proportional to the
neutrino mass matrix, one expects all possible flavour combinations of
the four leptons to occur, including lepton flavour violating ones. In
Ref.~\cite{Han:2007bk} simple cuts have been defined for final states
consisting of electrons and muons, eliminating essentially any
Standard Model background.

In general tau reconstruction is experimentally more difficult because
of the missing transverse energy from neutrinos. However, in the case
of interest enough kinematic constraints should be available to
identify also events involving taus. It turns out that the inclusion
of such events significantly increases the sensitivity for neutrino
parameters. Therefore, following Ref.~\cite{Hektor:2007uu}, we assume
that events where one of the four leptons is a tau can also be
reconstructed.\footnote{To be conservative we do not include events with more
than one tau, since already the inclusion of events with one tau
provides enough information for our purposes.}
This should be possible efficiently, despite the complications
involving the tau reconstruction, since the invariant mass is known
from decays without tau, which can be used as kinematic constraint for
events of the type $\ell^\pm \ell^\pm \, \ell^\mp \tau^\mp$ for $\ell
= e$ or $\mu$. Furthermore, one can adopt the assumption that the
neutrinos carrying away the missing energy are aligned with the tau.

In principle it is difficult to distinguish a primary electron or muon
from the ones originating from leptonic tau decays. Since here we are
interested in investigating the flavour structure of the decays,
leptonically decaying taus might be a ``background'' for the Higgs
decays into electrons and muons, and vice versa. However, due to the
energy carried away by the two neutrinos from the leptonic tau decay,
a cut on the invariant mass of the like-sign leptons should eliminate
such a confusion very efficiently. It is beyond the scope of this work
to perform a detailed simulation and event reconstruction study. The
above arguments suggest that our assumptions are suitable to estimate
the sensitivity of the Higgs decays for neutrino parameters by the
procedure outlined in the following.

We define as our five observables the number of
like-sign lepton pairs with the flavour combinations
\beq\label{eq:flavours}
x = (ee),(e\mu),(\mu\mu),(e\tau),(\mu\tau) \,.
\eeq
Note that these five branchings contain the full information, since
$\text{BR}_{\tau\tau}$, which we do not use explicitly, is fixed by
$\text{BR}_{\tau\tau} = 1 - \sum_x \text{BR}_x$. Taking into account
the number of occurrences of the combinations (\ref{eq:flavours}) in
four leptons where at most one tau is allowed, the number of events in
each channel is obtained as:
\beq \label{eq:events}
\begin{aligned}
&N_{ab} =  2 N_{2H} \, \epsilon \, \text{BR}_{ab} \sum_x \text{BR}_x
&&\quad\text{for}\quad (ab) = (ee),(e\mu),(\mu\mu) \,, \\
&N_{ab} =  2 N_{2H} \, \epsilon \, \text{BR}_{ab} 
(\BRee + \BRem + \BRmm)
&&\quad\text{for}\quad (ab) = (e\tau),(\mu\tau) \,,
\end{aligned}
\eeq 
where $N_{2H}$ is the total number of doubly charged scalar pairs
decaying into four leptons, and $\epsilon$ is the detection efficiency
for the four lepton events. For simplicity we assume here a flavour
independent efficiency. The branching ratios are given in
Eq.~(\ref{BR}).
To illustrate the sensitivity to neutrino parameters we will use
$\epsilon N_{2H} = 10^3$ or $\epsilon N_{2H} = 10^2$ events. For an
integrated luminosity of 100~fb$^{-1}$ at LHC these event numbers will
be roughly obtained for $\MH \simeq 350$~GeV and $\MH \simeq 600$~GeV,
respectively~\cite{Han:2007bk}.

To carry out the analysis we define a $\chi^2$ function from the
observables in Eq.~(\ref{eq:events}). For given $\epsilon N_{2H}$ they
depend only on neutrino parameters. We consider five continuous
parameters: the lightest neutrino mass $m_0$, $s_{13}$, the Dirac
phase $\delta$, and the two Majorana phases $\alpha_{12} = \alpha_1 -
\alpha_2$ and $\alpha_{32}= \alpha_3 - \alpha_2$, plus the discrete
parameter $h =$~NH or IH describing the mass ordering. The remaining
neutrino parameters, the two mass-squared differences and the mixing
angles $s_{12}$ and $s_{23}$, are fixed to their experimental best fit
values given in Eq.~(\ref{eq:osc-params}). The $\chi^2$ is constructed
as:
\beq\label{eq:chisq}
\begin{split}
\chi^2(m_0,s_{13}, \delta, \alpha_{12},\alpha_{32}, h) = 
\sum_{xy} V_x \, S^{-1}_{xy} \, V_y + 
\left(\frac{s_{13}^2}{\sigma_{s_{13}^2}}\right)^2
\qquad\text{with} \\
V_x = N_x^\mathrm{pred}
(m_0,s_{13},\delta,\alpha_{12},\alpha_{32}, h) - N_x^\mathrm{exp}
\end{split}
\eeq 
where $x$ and $y$ run over the five combinations given in
Eq.~(\ref{eq:flavours}). For the ``data'' $N_x^\mathrm{exp}$ we use
the prediction for $N_x$ at some assumed ``true values'' of the
parameters, \linebreak
$(m_0, s_{13}, \delta, \alpha_{12}, \alpha_{32},
h)^\mathrm{true}$. Then the statistical analysis tells us the ability
to reconstruct these true values from the data.
For the covariance matrix $S$ we assume the following form:
\beq
S_{xy} = N_x^\mathrm{exp} \delta_{xy} + 
         \sigma^2_\mathrm{norm} N_x^\mathrm{pred} N_y^\mathrm{pred} + 
         S^\mathrm{osc}_{xy} \,.
\eeq
It includes statistical errors, a fully correlated normalisation error
$\sigma_\mathrm{norm}$, and the uncertainty introduced from the errors
on the oscillation parameters $S^\mathrm{osc}$. The normalisation
error $\sigma_\mathrm{norm}$ arises from the uncertainty on the
luminosity and the efficiency. Moreover, the possibility that the
non-leptonic decays of $H^{++}$ of Eq.~(\ref{eq:decays}) might occur
at a sub-leading level and are not observed introduces an uncertainty
in the number of leptonic decays. We adopt a value of
$\sigma_\mathrm{norm} = 20\%$. We have checked that even an analysis
with free normalization (i.e., $\sigma_\mathrm{norm} \to \infty$)
leads to very similar results. This means that the information is
fully captured by the ratios of branchings.\footnote{This is true as
long as all branchings from Eq.~(\ref{eq:flavours}) are used; if the
events containing taus are omitted our results depend to some degree
on the value adopted for $\sigma_\mathrm{norm}$.}

Via the covariance matrix $S^\mathrm{osc}$ we account for the fact
that the parameters $\Delta m^2_{21}$, $|\Delta m^2_{31}|$, $s_{12}$
and $s_{23}$ have a finite uncertainty. We include the errors from
Eq.~(\ref{eq:osc-params}) and take into account the correlations which
they introduce between the observables $N_x$. 
The last term in Eq.~(\ref{eq:chisq}) takes into account the
constraint on $s_{13}$ from present data according to
Eq.~(\ref{eq:th13}). Let us note that within the time scale of a few
years the errors on oscillation parameters are likely to decrease.  In
particular, also the bound on $s_{13}$ will be strengthened or
eventually a finite value could be discovered by upcoming reactor or
accelerator experiments, see for example Ref.~\cite{Huber:2004ug}. To
be conservative we include only present information, although at the
time of the analysis better constraints might be available. We have
checked that the precise value of $s_{13}$ within the current limits
as well as its uncertainty have a very small impact on our results,
and a better determination may lead at most to a marginal
improvement of the sensitivities.

\subsection{Branching ratios}
\label{sec:branching}

\begin{figure}[t!]
\centering
\includegraphics[width=0.98\textwidth]{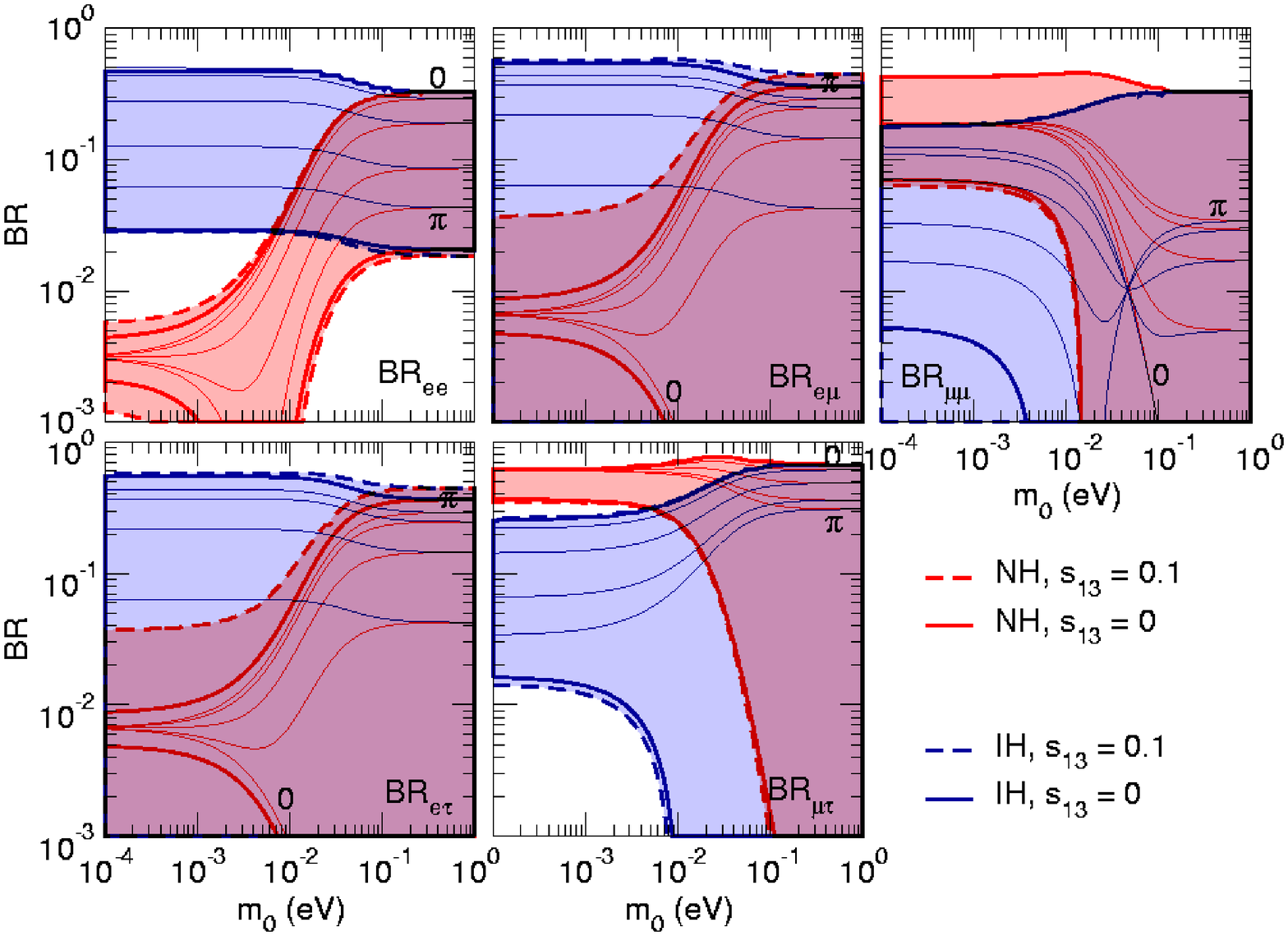}
  \mycaption{Branching ratios BR($H\to\ell_a\ell_b$) as function of
  the lightest neutrino mass $m_0$ for NH (light-red) and IH (dark-blue).  The
  thick solid lines are for $s_{13} = 0$, and the thick dashed lines
  for $s_{13}=0.1$, where the dependence on phases as well as the
  uncertainty of solar and atmospheric oscillation parameters at
  $2\sigma$ are included. The thin solid lines show the branchings for
  oscillation parameters fixed at the best fit points
  Eq.~(\ref{eq:osc-params}), $s_{13} = 0$, $\alpha_{32} = \pi$, and
  $\alpha_{12} = 0, \pi/4, \pi/2, 3\pi/4, \pi$.}
\label{fig:branching}
\end{figure}

In Fig.~\ref{fig:branching} we show the branching ratios for NH and IH
as a function of the lightest neutrino mass $m_0$. For fixed $m_0$,
the interval for the branching emerges due to the dependence on the
phases $\alpha_{12},\alpha_{32},\delta$, and also the uncertainty on
solar and atmospheric oscillation parameters contributes to the
interval. In the plots one can identify the regions of hierarchical
neutrino masses, $m_0 < 10^{-3}$~eV, and QD masses, $m_0 > 0.1$~eV,
where NH and IH become indistinguishable. In the limiting cases $m_0
= 0$ and $m_0 \to\infty$ the analytic expressions for the branchings
are rather simple. For NH and $m_0 = 0$ one finds to leading order in
the small quantities $r \equiv \Delta m^2_{21}/|\Delta m^2_{31}|
\approx 0.03$ and $s_{13}^2 < 0.05$ (at $3\sigma$):
\bea
  \BRee^{\mathrm{NH},m_0 = 0} &\approx& 
    s_{12}^4 r + 2 s_{12}^2 s_{13}^2 \sqrt{r} \cos(\alpha_{32} - 2\delta) \,,
    \label{eq:BReeNH} \\
  \BRem^{\mathrm{NH},m_0 = 0} &\approx& 
    2\left[ s_{12}^2c_{12}^2c_{23}^2 r + s_{23}^2 s_{13}^2 + 
            2 s_{12} c_{12} s_{23} c_{23} s_{13} \sqrt{r} \cos(\alpha_{32}-\delta)
    \right] \,,\label{eq:BRemNH}\\
  \BRmm^{\mathrm{NH},m_0 = 0} &\approx& 
    s_{23}^4 + 2 s_{23}^2 c_{23}^2 c_{12}^2 \sqrt{r} \cos\alpha_{32} + c_{23}^4 c_{12}^4 r 
    \nonumber\\
   &&\hspace*{2cm}
     - 4 s_{23}^3 c_{23} s_{12} c_{12} s_{13} \sqrt{r} \cos (\alpha_{32} - \delta) \,,
    \label{eq:BRmmNH}\\
  \BRet^{\mathrm{NH},m_0 = 0} &\approx& 
    2\left[ s_{12}^2 c_{12}^2 s_{23}^2 r + c_{23}^2 s_{13}^2 -
            2 s_{12} c_{12} s_{23} c_{23} s_{13} \sqrt{r} \cos(\alpha_{32}-\delta)
    \right] \,,\label{eq:BRetNH}\\
  \BRmt^{\mathrm{NH},m_0 = 0} &\approx& 
    2s_{23}^2c_{23}^2\left(1  - 2c_{12}^2\sqrt{r}\cos\alpha_{32} + c_{12}^4 r
   \right)\,.\label{eq:BRmtNH}
\eea
For IH and $m_0 = 0, s_{13} = 0$ we have
\bea
  \BRee^{\mathrm{IH},m_0 = 0} &=& 
    \frac{1}{2} 
    \left(1 - \sin^22\theta_{12}\sin^2\frac{\alpha_{12}}{2}\right) \,,
    \label{eq:BReeIH}\\
  \BRem^{\mathrm{IH},m_0 = 0} &=& 
    c_{23}^2 \sin^22\theta_{12}\sin^2\frac{\alpha_{12}}{2}\,,
    \label{eq:BRemIH}\\
  \BRmm^{\mathrm{IH},m_0 = 0} &=& 
    \frac{c_{23}^4}{2} 
    \left(1 - \sin^22\theta_{12}\sin^2\frac{\alpha_{12}}{2}\right)\,,
    \label{eq:BRmmIH}\\
  \BRet^{\mathrm{IH},m_0 = 0} &=& 
    s_{23}^2 \sin^22\theta_{12}\sin^2\frac{\alpha_{12}}{2}\,,
    \label{eq:BRetIH}\\
  \BRmt^{\mathrm{IH},m_0 = 0} &=& 
    \frac{1}{4}\sin^22\theta_{23}
    \left(1 - \sin^22\theta_{12}\sin^2\frac{\alpha_{12}}{2}\right)\,,
    \label{eq:BRmtIH}
\eea
and in the limit $m_0\to\infty$ with $s_{13}=0$ the branchings
become
\bea
\BRee^\mathrm{QD} &=& \frac{1}{3}
    \left(1 - \sin^22\theta_{12} \, \sin^2\frac{\alpha_{12}}{2}\right) 
    \, = \,\frac{2}{3}\, \BRee^{\mathrm{IH},m_0 = 0} \,,
     \label{eq:BReeQD}\\
\BRem^\mathrm{QD} &=& \frac{2}{3} \, c_{23}^2 \,
    \sin^22\theta_{12} \, \sin^2\frac{\alpha_{12}}{2} 
    \quad =\, \frac{2}{3} \, \BRem^{\mathrm{IH},m_0 = 0} \,,
     \label{eq:BRemQD}\\
\BRmm^\mathrm{QD} &=& \frac{1}{3}\left[
    1 - \frac{1}{2}\sin^22\theta_{23} 
      \left(1 - s_{12}^2\cos\alpha_{31} - c_{12}^2\cos\alpha_{32}\right)
      - c_{23}^4 \sin^22\theta_{12} \sin^2\frac{\alpha_{12}}{2}
    \right]\,,
    \label{eq:BRmmQD}\\
\BRet^\mathrm{QD} &=& \frac{2}{3} \, s_{23}^2 \,
    \sin^22\theta_{12} \, \sin^2\frac{\alpha_{12}}{2} 
    \, = \,\frac{2}{3} \, \BRet^{\mathrm{IH},m_0 = 0} \,,
     \label{eq:BRetQD}\\
\BRmt^\mathrm{QD} &=& \frac{1}{3}\sin^22\theta_{23}\left(
    1 - s_{12}^2\cos\alpha_{31} - c_{12}^2\cos\alpha_{32}
      - \frac{1}{2}\sin^22\theta_{12} \sin^2\frac{\alpha_{12}}{2}
    \right)\,.
    \label{eq:BRmtQD}
\eea
Note that for a vanishing lightest neutrino mass, $m_0=0$, there is
only one physical Majorana phase, $\alpha_{32}$ for NH, and
$\alpha_{12}$ for IH, as clear from Eqs.~(\ref{eq:diag}) and
(\ref{eq:U}).

In the following we will explore the parameter dependencies of these
branchings to obtain information on the neutrino mass spectrum and on
Majorana phases. The rather wide ranges for the branchings in the
cases of IH and QD spectrum suggest a strong dependence on the phases,
and as we will see in Sec.~\ref{sec:phases} these are the cases where
Majorana phases can be measured very efficiently. The determination of
the mass spectrum is somewhat more subtle.

A clear signature for the NH with small $m_0$ is provided by
\BRee\footnote{Note that the behaviour of \BRee\ is the same as the
effective neutrino mass probed in neutrino-less double beta-decay,
which is also proportional to $|M_{ee}|$, see for example
Ref.~\cite{Pascoli:2005zb}.}. Eq.~(\ref{eq:BReeNH}) shows that for NH
and $m_0 = 0$, \BRee\ is suppressed by $r$ and/or $s_{13}^2$, and
there is the upper bound $\BRee < 5.3\times 10^{-3}$ for the largest
value of $s^2_{12}$ allowed at $2\sigma$ and $s_{13}^2 = 0.01$, in
agreement with Fig.~\ref{fig:branching}. In contrast, for IH with $m_0
< 0.01$~eV and for QD spectrum, Eqs.~(\ref{eq:BReeIH}) and
(\ref{eq:BReeQD}) give the lower bounds $\BRee > (1-
\sin^22\theta_{12})/2 \approx 0.03$ and $\BRee > (1-
\sin^22\theta_{12})/3 \approx 0.02$, respectively. Therefore, the
characteristic signature of normal hierarchical spectrum is the
suppression of Higgs decays into two electrons.

From a first glance at Fig.~\ref{fig:branching} one could expect that
it might be difficult to distinguish IH and QD spectra, since there is
always overlap between the allowed regions in the branchings. Indeed,
if only branchings involving electrons and muons (\BRee, \BRem, \BRmm)
are considered there is some degeneracy between IH and QD, especially
if $s_{13}$ is allowed to be close to the present bound. However, as
we will show, due to the complementary dependence on the phases of 
all the BR$_{ab}$ including also taus, the degeneracy is broken and
these two cases can be disentangled. Consider, for example, \BRmm\ and
\BRmt: in the case of IH with $m_0=0$ they behave very similar as a
function of $\alpha_{12}$, see Eqs.~(\ref{eq:BRmmIH}) and
(\ref{eq:BRmtIH}), whereas for QD they show opposite dependence,
compare Eqs.~(\ref{eq:BRmmQD}) and (\ref{eq:BRmtQD}), and phases which
give $\BRmm^\mathrm{QD} = 0$ maximise $\BRmt^\mathrm{QD}$.

Note that for $s_{13}=0$ and $s_{23}^2 = 0.5$, \BRem\ and \BRet\ are
identical. Nevertheless there is important complementariness between
them. First, the uncertainty on $s_{23}^2$, see
Eq.~(\ref{eq:osc-params}), affects each of them significantly, and it
reduces the final sensitivity if only \BRem\ is used in the analysis.
But since \BRem\ and \BRet\ are related by the transformation $s_{23}
\to c_{23}$, $c_{23} \to -s_{23}$ this uncertainty is cancelled if both
of them are included in the fit. Second, it can be shown that the
leading order term in $s_{13}$ is the same for \BRem\ and \BRet,
apart from an opposite sign. Therefore, also the impact of $s_{13}$ is
strongly reduced if information from both of them is taken into
account.
One can observe from Fig.~\ref{fig:branching} that for small $m_0$ and
NH, \BRem\ and \BRet\ show a significant dependence on $s_{13}$, while
in the other cases the dependence is mild. The reason is a leading
term linear in $\sqrt{r} s_{13}$ in Eqs.~(\ref{eq:BRemNH}) and
(\ref{eq:BRetNH}), whereas in all other cases $s_{13}$ appears either
in sub-leading terms or at least at second order.

\subsection{Determination of the neutrino mass spectrum}
\label{sec:spectrum}

\begin{figure}
\centering
\includegraphics[width=0.98\textwidth]{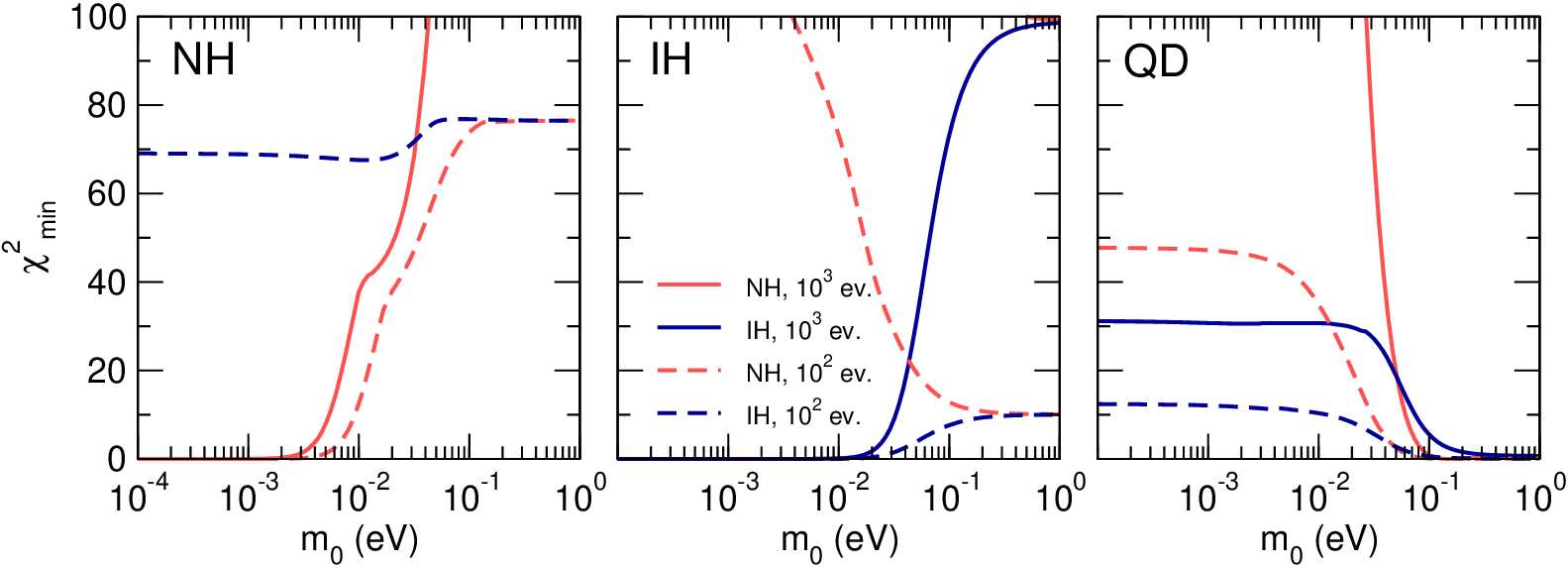}
  \mycaption{$\chi^2_\mathrm{min}$ vs $m_0$ assuming a true
  hierarchical spectrum with NH (left) and IH (middle), and a true QD
  spectrum (right). The $\chi^2$ is shown for $\epsilon N_{2H} = 100$
  (dashed) and 1000 (solid) events, and $\sigma_\mathrm{norm} =
  20\%$. In the fit we assume either NH (light-red) or IH (dark-blue),
  and we minimise with respect to $s_{13}$ and the phases. We adopt
  the following true parameter values. Left: $m_0 = 0$, NH,
  $\alpha_{32} = \pi$; middle: $m_0 = 0$, IH, $\alpha_{12} = 0$;
  right: $m_0 = 0.15$~eV, $\alpha_{12} = 0.1\pi$, $\alpha_{32} =
  1.6\pi$; and always $s_{13}=0$.}
  \label{fig:mass-spectrum}
\end{figure}

Let us now quantify the ability to determine the neutrino mass
spectrum by performing a $\chi^2$ analysis as described in
Sec.~\ref{sec:analysis}.  In Fig.~\ref{fig:mass-spectrum} we show the
$\chi^2$ by assuming that ``data'' are generated by a hierarchical
spectrum with normal ordering (left), a hierarchical spectrum with
inverted ordering (middle), or a QD spectrum (right). These data are
fitted with both possibilities for the ordering (NH, light-red curves, and
IH, dark-blue curves) and a value for $m_0$ shown on the horizontal
axis. We minimise the $\chi^2$ with respect to the other parameters,
taking into account the current bound on $s_{13}$. The results are
shown for a total number of doubly charged scalars decaying into
like-sign leptons of $\epsilon N_{2H} = 10^3$ (solid) and $10^2$
(dashed).

First we discuss the sensitivity to hierarchical spectra with a
very small lightest neutrino mass $m_0$. The left panel of
Fig.~\ref{fig:mass-spectrum} shows that a NH with small $m_0$ can be
identified with very high significance. An inverted hierarchical
spectrum as well as a QD spectrum have $\Delta\chi^2 \gtrsim 60$
already for 100 events. An upper bound on the lightest neutrino mass
of $m_0 \lesssim 0.01$~eV at $3\sigma$ can be established by LHC
data. As discussed in the previous section this information comes
mainly from the suppression of the decay into two electrons, which
occurs only for normal hierarchical spectrum. An inverted hierarchical
spectrum (middle panel) can be distinguished from a QD one at around
3$\sigma$ with 100 events, where the $\chi^2$ increases roughly
linearly with the number of events. The ability to exclude a QD
spectrum in case of a true IH depends on the true value of the
Majorana phase $\alpha_{12}$. The example chosen in
Fig.~\ref{fig:mass-spectrum}, $\alpha_{12}^\mathrm{true} = 0$,
corresponds to the worst case; for all other values of $\alpha_{12}$
the $\chi^2$ for QD is bigger.

\begin{figure}[t]
\centering
 \includegraphics[width=0.8\textwidth]{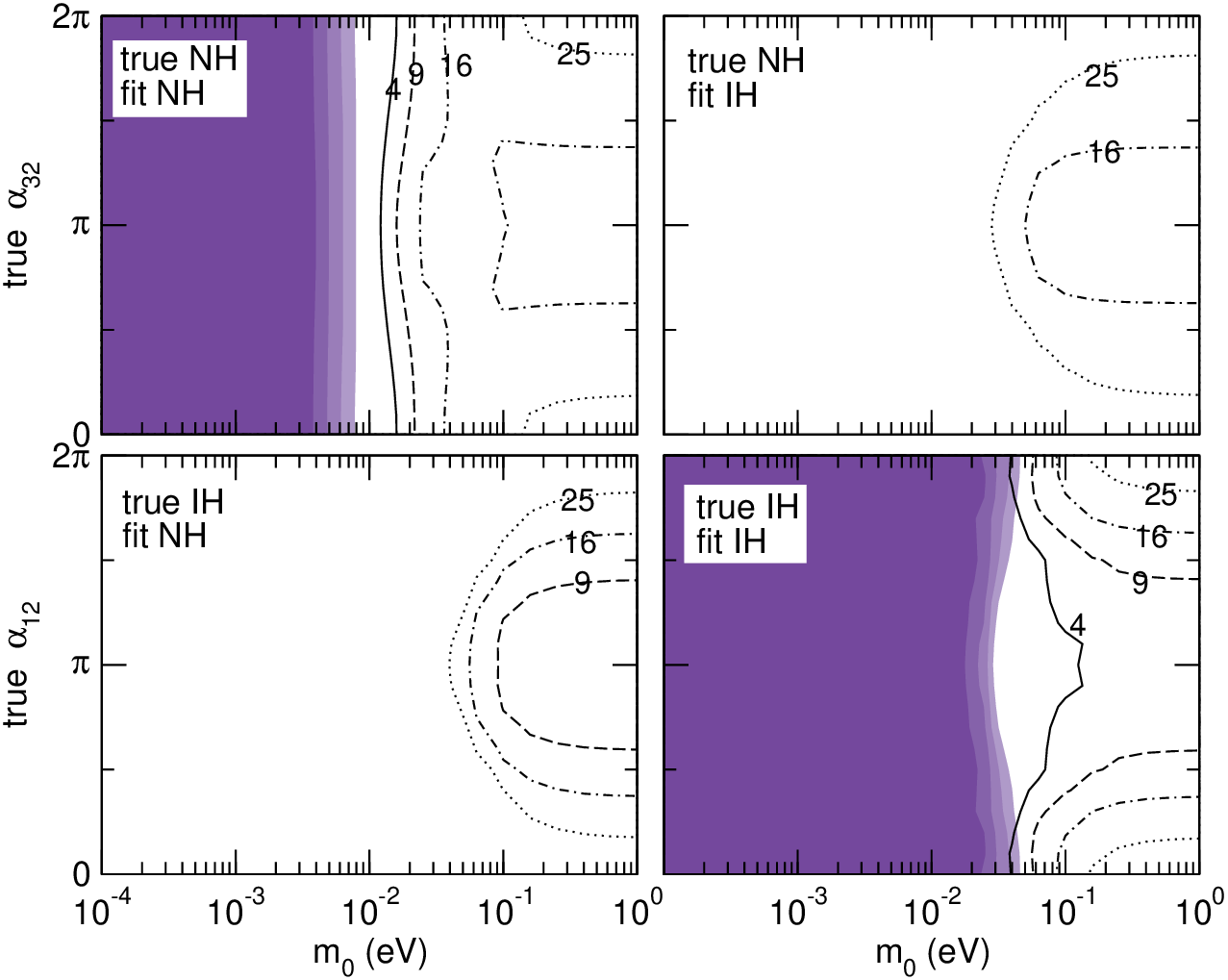}
  \mycaption{Determination of hierarchical neutrino mass spectra,
  $m_0^\mathrm{true} = 0$, assuming 1000 Higgs pair decays. The upper
  (lower) panels are for a true NH (IH), and for the left (right)
  panels the fit is performed assuming a NH (IH). As a function of the
  true value of the Majorana phases we show contours $\chi^2 =
  4,9,16,25$ (from dark to light), minimising with respect to all
  parameters except from $m_0$. Coloured regions correspond to our
  standard analysis, whereas for the black contours we do not use
  decays into tau leptons.}
  \label{fig:spect-hierarchical}
\end{figure}

Fig.~\ref{fig:spect-hierarchical} shows the ability to identify a
hierarchical spectrum as a function of the true value for the Majorana
phase, where for $m_0=0$ there is only one physical phase. The shaded
regions show that for 1000 events the true spectrum can be identified
at 5$\sigma$ significance, and an upper bound on the lightest neutrino
mass $m_0 < 8\times 10^{-3}$~eV for NH and $m_0 < 4\times 10^{-2}$~eV
for IH is obtained, independent of the true phase. For the black
contours in Fig.~\ref{fig:spect-hierarchical} we do not use the
information from decays into taus, i.e., we use only the lepton pairs
$(ee), (e\mu), (\mu\mu)$. This analysis illustrates the importance of
the tau events. For example, if tau events are not used an IH with
$m_0=0$ cannot be distinguished from a QD spectrum for
$\alpha_{12}^\mathrm{true} \sim \pi$. Also the sensitivity to a NH is
significantly reduced, which becomes even more severe if less events
were available.

\begin{figure}[t]
\centering
 \includegraphics[width=0.9\textwidth]{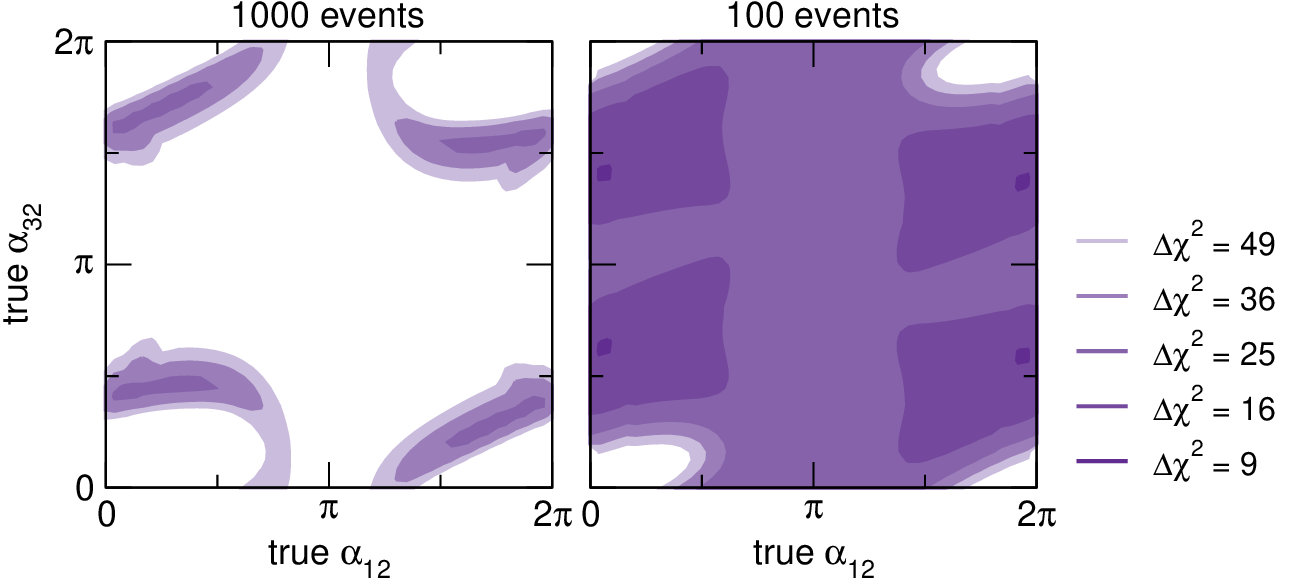}
  \mycaption{Exclusion of an IH with $m_0 = 0$ in the case of a true
  QD spectrum. We show $\chi^2$ contours for 1000 events (left) and
  100 events (right) in the plane of the true Majorana phases
  assuming a true QD spectrum ($m_0^\mathrm{true} = 0.15$~eV,
  $s_{13}^\mathrm{true} = 0$) fitted with IH and $m_0 = 0$, minimising
  with respect to all other parameters.}
  \label{fig:spectrum-QD}
\end{figure}

Now we move to the discussion of a true QD spectrum. As shown in the
right panel of Fig.~\ref{fig:mass-spectrum} also a QD spectrum can be
identified quite well, and a lower bound on the lightest neutrino mass
of $m_0 > 2\,(6)\times 10^{-2}$~eV at $3\sigma$ can be obtained for
100 (1000) events. Note that for the example shown in
Fig.~\ref{fig:mass-spectrum}, 100 events give a $\Delta\chi^2 \approx
12.4$ for the IH with $m_0=0$, which corresponds roughly to an
exclusion at $3.5\sigma$. The potential to exclude a hierarchical
inverted spectrum depends on the true values of the Majorana phases,
and the true values of $\alpha_{12}$ and $\alpha_{32}$ adopted in
Fig.~\ref{fig:mass-spectrum} correspond to the worst sensitivity.
In Fig.~\ref{fig:spectrum-QD} we show contours of $\Delta\chi^2$ for
IH with $m_0=0$ assuming a true QD spectrum, in the plane of the true
Majorana phases. For 1000 events we find some islands in the plane of
$\alpha_{12}$ and $\alpha_{32}$ where the $\chi^2$ reaches values as
low as 30 (compare Fig.~\ref{fig:mass-spectrum}), however in most
parts of the parameter space the exclusion is at more than $7\sigma$.
For 100 events typically a significance better than $4\sigma$ is
reached, but there are some notable regions ($-\pi/2 \lesssim
\alpha_{12} \lesssim \pi/2$ and $\alpha_{32} \sim \pi/2, 3\pi/2$)
with $\chi^2$ values between 16 and 9.

Let us add that for the exclusion of an inverted hierarchical spectrum
in the case of a true QD spectrum the branchings into tau leptons are
crucial. If only electron and muon events are used in most regions of
the parameter space an IH with $m_0=0$ can fit data from a QD
spectrum. For $(ee),(e\mu),(\mu\mu)$ branchings a degeneracy between
IH and QD appears due to the freedom in adjusting $s_{13}$, $\delta$,
$\theta_{23}$ and the Majorana phases. This effect is also apparent
from the black contour lines in Fig.~\ref{fig:spect-hierarchical}
(lower-right panel).  The significance of this degeneracy depends on
details such as the errors imposed on $s_{13}^2$ and $s_{23}^2$, as
well as on the systematical error $\sigma_\mathrm{norm}$. As discussed
in Sec.~\ref{sec:branching}, taking into account also decays into
$e\tau$ and $\mu\tau$ is crucial to break this degeneracy, and in the
full analysis used to calculate Figs.~\ref{fig:mass-spectrum} and
\ref{fig:spectrum-QD} the dependence on subtleties such as $s_{13}$
and $\sigma_\mathrm{norm}$ is small.

\subsection{Determination of Majorana phases}
\label{sec:phases}

Let us now investigate the tantalising possibility to determine the
Majorana phases $\alpha_{ij} \equiv \alpha_i - \alpha_j$ from the
doubly charged Higgs decays. Since the decay is governed by a single
diagram without any interference term the decays are CP conserving,
and therefore no explicit CP violating effects can be observed.
Nevertheless, the branchings depend (in a CP conserving way) on the
phases, which eventually may allow to establish CP violating values
for them. 
In general the measurement of Majorana phases is a very difficult
task. Probably the only hope to access these phases will be
neutrino-less double beta-decay in combination with an independent
neutrino mass determination, where under very favourable
circumstances~\cite{Pascoli:2005zb} the phase $\alpha_{12}$ might be
measurable.

We start by discussing some general properties of the branchings
related to the Majorana phases. Using Eqs.~(\ref{eq:diag}) and
(\ref{eq:U}) one can write:
\beq\label{eq:BR-CP}
\text{BR}_{ab} \propto |M_{ab}|^2 =
  \left| \sum_{i=1}^3 V_{ai} V_{bi} \, e^{i\alpha_i} \, m_i \right|^2 \,,
\eeq
From this expression it is evident that for a vanishing lightest
neutrino mass, $m_0=0$, there is only one physical Majorana phase,
$\alpha_{32}$ for NH and $\alpha_{12}$ for IH. Next we note that since
$V_{e3} \propto s_{13}$, it is clear that for $s_{13} = 0$ all
branchings involving electrons can only depend on
$\alpha_{12}$.\footnote{For the same reason only $\alpha_{12}$ can be
tested in neutrino-less double beta-decay, where $|M_{ee}|$ is
probed.}  Since the small effects of $s_{13}$ cannot be explored
efficiently, the determination of both phases simultaneously
necessarily involves \BRmm\ and/or \BRmt, see also
Eqs.~(\ref{eq:BReeNH}) to (\ref{eq:BRmtQD}).  Furthermore, from
Eq.~(\ref{eq:BR-CP}) it can be seen that the branchings are invariant
under
\beq\label{eq:symmetry}
\alpha_{ij} \to 2\pi - \alpha_{ij} \,,\qquad \delta \to 2\pi - \delta \,.
\eeq
This symmetry is a consequence of the fact that there is no CP
violation in the decays, and therefore the branchings have to be
invariant under changing the signs of all phases simultaneously.

\begin{figure}
\centering
\includegraphics[width=0.98\textwidth]{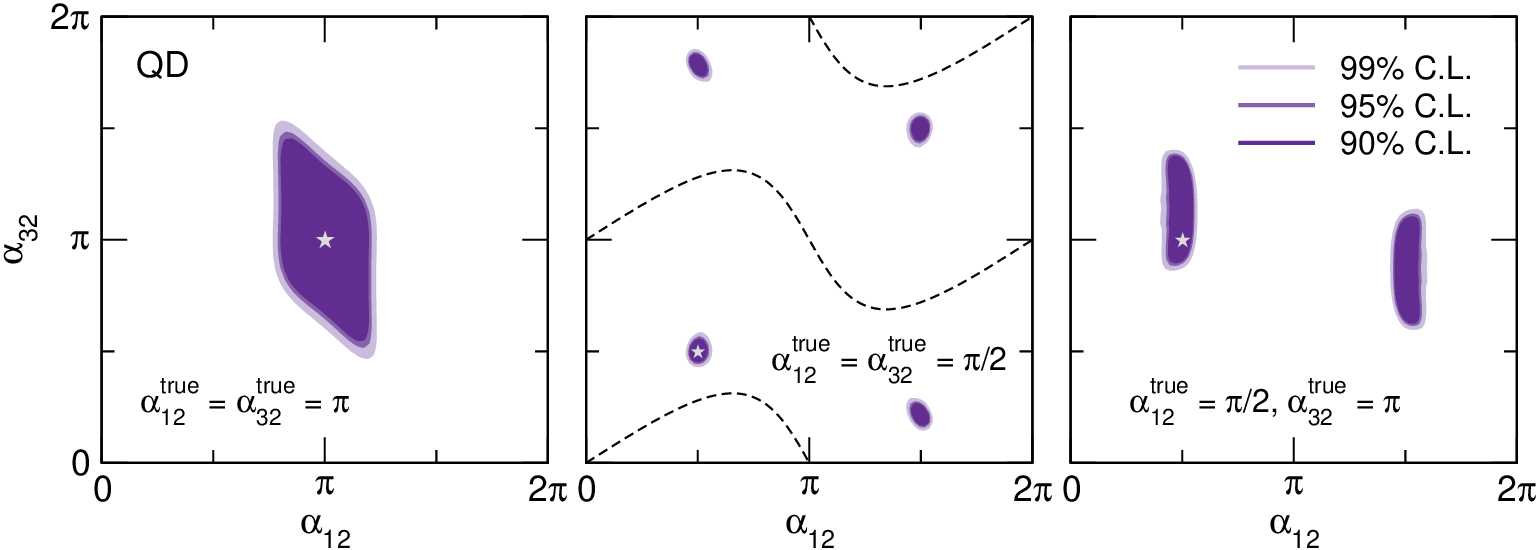}
  \mycaption{Determination of the Majorana phases for QD spectrum
  ($m_0 = 0.15$~eV) from 1000 doubly-charged Higgs pair events. We
  assume $s_{13}^\mathrm{true} = 0$ and three example points for the
  true values of the Majorana phases given in each panel. The dashed
  lines in the middle panel correspond to the true values of the
  phases for which the degenerate solution according to
  Eq.~(\ref{eq:deg-a32}) appears at a CP conserving value of
  $\alpha_{32}$.}
  \label{fig:phases-QD}
\end{figure}

In Fig.~\ref{fig:phases-QD} we show that for a QD spectrum the
observation of the decay of 1000 doubly-charged Higgs pairs allows to
determine both Majorana phases. We assume some true values for the two
phases and then perform a fit leaving all parameters free, where for
$s_{13}$ we impose the constraint from present data. The actual
accuracy to determine the phases depends on their true values, where
we show three different examples in the three panels. For
$\alpha_{12}=\alpha_{32}=\pi$ (left panel) the allowed region is the
largest, however the phases can be constrained to a unique region. In
the other two cases the accuracy is better, but some ambiguities are
left. The symmetry from Eq.~(\ref{eq:symmetry}) is apparent in all
panels, whereas in the case $\alpha_{12}=\alpha_{32}=\pi$ it does not
introduce an ambiguity.

The features of Fig.~\ref{fig:phases-QD} can be understood from
Eqs.~(\ref{eq:BReeQD}) to (\ref{eq:BRmtQD}). In addition to the
symmetry Eq.~(\ref{eq:symmetry}) one finds that in the limit
$s_{13}=0$ the phases $\alpha_{31}$ and $\alpha_{32}$ appear only in
the particular combination
\beq\label{eq:comb-a32} 
(s_{12}^2\cos\alpha_{31} + c_{12}^2\cos\alpha_{32}) \propto 
\cos(\alpha_{32} - \varphi)
\qquad\text{with}\qquad
\tan\varphi = \frac{s_{12}^2 \sin\alpha_{12}}
{c_{12}^2 + s_{12}^2 \cos\alpha_{12}} \,,
\eeq
where we have used $\alpha_{12}$ and $\alpha_{32}$ as independent
parameters. For constant $\alpha_{12}$ there are two values of
$\alpha_{32}$ which leave this combination invariant: for each
$\alpha_{32}$ we expect a degenerate solution at
\beq\label{eq:deg-a32}
\alpha_{32}' = 2\varphi - \alpha_{32} \,.
\eeq
For $\alpha_{12} = \pi/2$ one finds $2\varphi \approx 0.28\pi$.  In
the case of $\alpha_{32} = \pi$ shown in the right panel of
Fig.~\ref{fig:phases-QD} this degenerate solution appears at
$\alpha_{32}' \simeq 1.28\pi$, which cannot be resolved from the
original one, and we are left with a two-fold ambiguity, due to
Eq.~(\ref{eq:symmetry}). In the middle panel, for $\alpha_{12} =
\alpha_{32} = \pi/2$, the ambiguity (\ref{eq:deg-a32}) leads to a
separated solution around $\alpha_{32}' \simeq 1.78\pi$ and, together
with the symmetry from Eq.~(\ref{eq:symmetry}) we end up with four
degenerate solutions. However, in this case the individual regions are
rather small, and the CP violating values of both phases can be
established despite the presence of the four-fold ambiguity.

\begin{figure}
\centering
\includegraphics[width=0.8\textwidth]{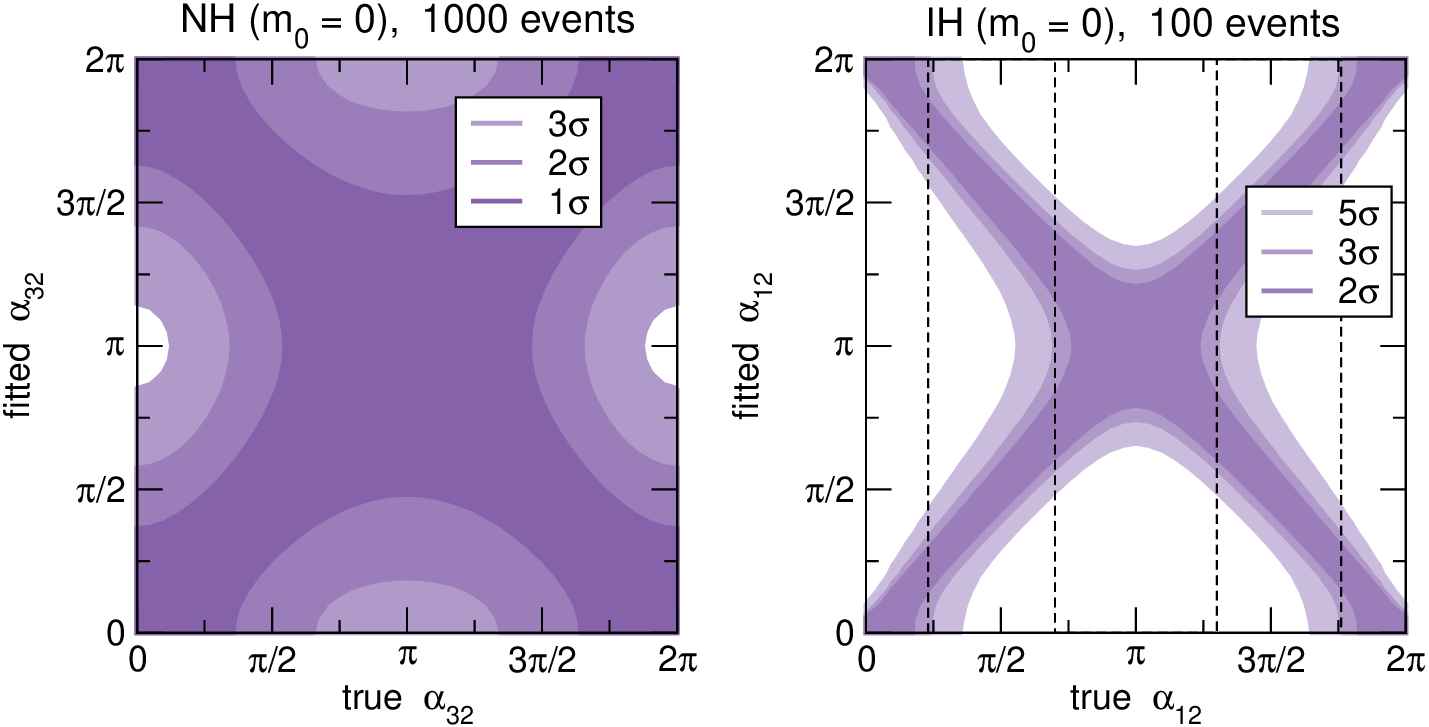}
  \mycaption{Determination of the Majorana phase for vanishing
  lightest neutrino mass. We assume $s_{13}^\mathrm{true} = 0$. Left:
  1, 2, 3$\sigma$ ranges for $\alpha_{32}$ as a function of its true
  value for NH assuming 1000 doubly-charged Higgs pair events. Right:
  2, 3, 5$\sigma$ ranges for $\alpha_{12}$ as a function of its true
  value for IH assuming 100 doubly-charged Higgs pair events. The
  dashed vertical lines indicate the region where CP violating values
  of $\alpha_{12}$ can be established at 3$\sigma$.}
  \label{fig:phases-NH-IH}
\end{figure}

Note that the symmetry (\ref{eq:symmetry}) does not mix CP
conserving and violating values of the phases, whereas this can happen
for the degeneracy Eq.~(\ref{eq:deg-a32}). The dashed curves in the
middle panel of Fig.~\ref{fig:phases-QD} correspond to the true
values of the phases, for which $\alpha_{32}' = 0$ or $\pi$. Hence,
along these curves CP violating values for $\alpha_{32}$ cannot be
established since the degeneracy is located at a CP conserving value.

Let us now discuss the potential to determine Majorana phases in case
of hierarchical spectra. As mentioned above, in this case there is
only one physical phase, $\alpha_{32}$ for NH and $\alpha_{12}$ for
IH. In Fig.~\ref{fig:phases-NH-IH} we show the allowed interval for
this phase which is obtained from the data as a function of its true
value.  In the fit the $\chi^2$ is minimised with respect to all other
parameters.  The left panel shows that for NH even with 1000 events at
most a 2$\sigma$ indication can be obtained, on whether $\alpha_{32}$
is closer to zero or $\pi$. This can be understood from
Eqs.~(\ref{eq:BReeNH}) to (\ref{eq:BRmtNH}), which show that
$\alpha_{32}$ appears at least suppressed by $\sqrt{r}$.  In
contrast, as visible in the right panel, for IH a rather precise
determination of $\alpha_{12}$ is possible already for 100 events,
apart from the ambiguity $\alpha_{12} \to 2\pi - \alpha_{12}$. For
$\alpha_{12}$ around $\pi/2$ or $3\pi/2$ its CP violating value can be
established, as marked by the vertical lines in
Fig.~\ref{fig:phases-NH-IH}. The good sensitivity is obvious from
Eqs.~(\ref{eq:BReeIH}) to (\ref{eq:BRmtIH}), which show a strong
dependence of the leading terms in the branchings on $\alpha_{12}$.

\section{Summary and concluding remarks}
\label{sec:conclusions}

In this work we have adopted the assumptions that ({\it i}) neutrino
masses are generated by the VEV of a Higgs triplet, ({\it ii}) the
doubly charged component of the triplet is light enough to be
discovered at LHC, i.e., lighter than about 1~TeV, and ({\it iii}) it
decays with a significant fraction into like-sign lepton pairs. We
have shown that under these assumptions LHC will provide very
interesting information for neutrino physics. The reason is that the
branching ratio of the doubly charged Higgs into like-sign leptons of
flavour $a$ and $b$, BR($H^{++}\to \ell^+_a \ell^+_b$), is
proportional to the modulus of the corresponding element of the
neutrino mass matrix, $|M_{ab}|^2$. Hence the flavour composition of
like-sign lepton events at LHC provides a direct test of the neutrino
mass matrix.

We have shown that the type of the neutrino mass spectrum (normal
hierarchical, inverted hierarchical, or quasi-degenerate) can be
identified at the $3\sigma$ level already with 100 doubly charged
Higgs pairs $H^{--}H^{++}$ decaying into four leptons. Typically such
a number of events will be achieved for doubly charged scalar masses
below 600~GeV and 100~fb$^{-1}$ integrated luminosity, whereas for
masses of 350~GeV of order 1000 events will be obtained. We have found
that it is possible to decide whether the lightest neutrino mass is
smaller or larger than roughly 0.01~eV, which marks the transition
between hierarchical and quasi-degenerate spectra. If it is smaller
the mass ordering (normal vs inverted) can be identified. A
hierarchical spectrum with normal ordering has a distinct signature,
namely a very small branching of the doubly charged Higgs decays into
two electrons. Therefore, this mass pattern can easily be confirmed or
ruled out at very high significance level. The other two possibilities
for the neutrino mass spectrum, inverted hierarchical or
quasi-degenerate, are somewhat more difficult to distinguish, but also
in this case very good sensitivity is obtained, depending on the
observed number of events.

In this respect the inclusion of final states involving tau leptons is
important, since if only electrons and muons are considered a
degeneracy between IH and QD spectra appears. In our analysis we have
conservatively assumed that events where one of the four charged
leptons is a tau can be reconstructed efficiently, thanks to the
kinematic constraints and the information on the invariant mass of
the event available from events without a tau. Certainly a more
realistic study including detailed simulations and event
reconstruction should confirm the assumptions which we have adopted
here.

The decay of the doubly charged Higgs in this framework does not show
explicit CP violation, since the decay is dominated by a tree-level
diagram without any interference term which could induce CP
violation. Nevertheless, the CP conserving branching ratios strongly
depend on the Majorana CP phases of the lepton mixing
matrix. Therefore, the framework considered here opens the fascinating
possibility to measure the Majorana phases in the neutrino mass matrix
via CP even observables. Our results show that for an inverted
hierarchical spectrum as well as for quasi-degenerate neutrinos this
is indeed possible.  In the first case, there is only one physical
phase, $\alpha_{12}$, which can be determined up to an ambiguity
$\alpha_{12} \leftrightarrow 2\pi - \alpha_{12}$ already with 100
events. In the case of a quasi-degenerate spectrum both Majorana
phases can be measured, where, depending on the actual values some
ambiguities might occur. In many cases CP violating values of the
phases can be established.

Certainly the observation of a doubly charged scalar at LHC would be a
great discovery of physics beyond the Standard Model. Of course this
alone does by no means confirm the Higgs triplet mechanism for
neutrino masses, since doubly charged particles decaying into leptons
are predicted in many models. Therefore, in case such a particle is
indeed found at LHC various consistency checks will have to be
performed.
It might turn out that the relation BR$(H^{++}\to \ell^+_a \ell^+_b)
\propto |M_{ab}|^2$ cannot be fulfilled for any neutrino mass matrix
consistent with oscillation data. This would signal that a Higgs
triplet cannot be the only source for neutrino masses.  In this
respect the information from decays into leptons of all flavours
(including taus) will be important. For example, also in the Zee--Babu
model~\cite{Zee:1985id} for neutrino masses doubly charged scalars
might be found at LHC. However, in this case branchings into tau
leptons are suppressed by powers of $(m_\mu/m_\tau)^2$ with respect to
muons~\cite{Babu:2002uu}, whereas in the Higgs triplet model they are
of similar size because of close to maximal $\theta_{23}$ mixing.

If LHC data on BR$(H^{++}\to \ell^+_a \ell^+_b)$ will be consistent
with a neutrino mass matrix from oscillation data, an analysis as
pointed out in this work can be performed. Also in this case it will
be of crucial importance to cross check the results with independent
measurements, for example the determination of the neutrino mass
ordering by oscillation experiments, or the measurement of the
absolute neutrino mass in tritium beta-decay, neutrino-less double
beta-decay or through cosmological observations. In particular,
neutrino-less double beta-decay will provide a crucial test, since it
gives an independent determination of the $|M_{ee}|$ element of the
neutrino mass matrix, which---combined with information from
oscillation experiments---will further constrain the allowed flavour
structure of the di-lepton events at LHC. The next generation of
neutrino-less double beta-decay experiments is expected to probe the
regime of the QD neutrino spectrum within a timescale comparable to
the LHC measurement.
Information from searches for lepton flavour violating processes may
be used as additional important consistency checks for the model.

In conclusion, a TeV scale Higgs triplet offers an appealing mechanism
to provide mass to neutrinos, which can be directly tested at the 
LHC. Such a scenario opens the possibility to measure the Majorana
phases of the lepton mixing matrix, which in general is a very
difficult---if not a hopeless task.

\subsection*{Acknowledgement}

We thank Georges Azuelos, Kamal Benslama, Christophe Grojean, Tao Han,
Bob McElrath and Jure Zupan for very useful discussions. 
J.G. thanks CERN Theory Division for
hospitality during her stay and MEC for a FPU grant. 
This work was partially supported by MEC
Spanish grant FPA2007-60323.

\renewcommand{\baselinestretch}{1}


\end{document}